\documentstyle[aps,manuscript]{revtex}
\begin{document}
\title{Gamma Rays from Super-Heavy Relic Particles in the Halo}
\author{Pasquale Blasi}
\address{Department of Astronomy \& Astrophysics, and\\
Enrico Fermi Institute, The University of Chicago,\\
5640 South Ellis Avenue, Chicago, IL 60637}

\maketitle

\begin{abstract}
Superheavy (SH) quasistable particles generated in the Early 
Universe could be responsable for 
Ultra High Energy Cosmic Rays (UHECR) and be a component of
Cold Dark Matter (CDM) in the universe. These particles are likely to 
cluster in the galactic halo, so that the main part of UHECR are 
gamma rays produced in the decay of neutral pions. Charged pions
are also produced in the same process and result in high energy electrons.
We consider here the production of gamma rays by synchrotron emission of 
these electrons in the galactic magnetic field.
The gamma ray fluxes are in the region of interest for some current and
proposed experiments (e.g. EGRET, GLAST, MILAGRO) in the 
energy range $0.1-10^4$ GeV. A comparison with the existing upper limits 
at $10^5-10^8$ GeV is also carried out. The detection of this flux of 
gamma rays would be an important signature of
SH relic particles as sources of
UHECR and would give a clue to the physics of the Early Universe.

\end{abstract}


\section{Introduction}\label{intro}

The detection of cosmic rays well above $10^{20}$ eV poses a serious
challenge to our understanding of the acceleration processes as well
as of the propagation of these particles from the sources. Particles 
with $E\geq 10^{19}$ eV are thought to be of extragalactic origin
since the galactic magnetic field is unable to confine and isotropize
them. However if the sources are cosmological then the UHECR spectrum 
would have as a unique signature the presence of a pronounced cutoff 
(the so called GZK cutoff \cite{gzk}) at $E\sim 5\cdot 10^{19}$ eV, which is 
not observed (see \cite{agasa} for a recent discussion). This limits the 
distance to the sources of UHECR to be less than $\sim 50$ Mpc and no
plausible source has been identified by current experiments within 
this distance, unless 
very large deflection angles are assumed, which looks incompatible with 
the current limits on the intergalactic magnetic field (see \cite{kronberg}
for a review and \cite{bo} for revised limits in a inhomogeneous
universe).\par
The lack of counterparts to the observed events, together with the
difficulty of the known acceleration mechanisms to reach the observed 
energies, prompted a new class of models (top-down models) (for 
a recent review see \cite{sigl} and references therein) where UHECR
are produced by the decay of grand unification massive bosons (generically
named X-particles) with mass $m_X\sim 10^{13}-10^{16}$ GeV. 
These X-particles can either 
be produced in processes involving topological defects (TD) or 
by the decay of SH quasistable (lifetime $\tau$ much larger than
the age of the Universe $t_0$) particles relics of the early universe 
\cite{bkv,sarkar,kuz,bere,kolb1} in the assumption that they represent some  
fraction of the CDM.\par 
In the latter case, the UHECR spectrum is dominated by the halo 
component and hence it has no GZK cutoff. Signatures of this model have
been discussed in \cite{bbv}, where it was pointed out that if UHECR above
$\sim 10^{19}$ eV are produced in the halo by the decay of X-particles,
then the flux should be dominated by gamma rays produced by the decay 
of neutral pions from the hadronic cascade initiated by ultra high 
energy partons ($X\to q \bar q$, $q\to $ hadrons). The model is also 
characterized by some degree of anisotropy due to the asymmetric position
of the sun in the galaxy \cite{russi,bbv,bm}.\par
Together with neutral pions charged pions are also generated which in turn
decay into electrons and positrons (hereafter we shall call them both
electrons) and neutrinos.\par
In this paper we calculate the flux of gamma rays which is produced by 
synchrotron emission of the electrons in the galactic magnetic field
when the density of SH particles in the halo is normalized to fit the
observed UHECR flux. Our calculations show that in most of the
cases the predicted gamma ray 
emission is compatible with the upper limits imposed by experiments
like HEGRA, EASTOP, CASA-MIA and Utah-Michigan at $10^5-10^8$ GeV and,
in the energy range $0.1-10^4$ GeV might be detectable by
future experiments like GLAST.\par

The paper is planned as follows: in section \ref{simple} we present a benchmark
calculation where we estimate the magnitude of the gamma ray flux from
ultra high energy electrons (UHEE) in the galactic magnetic field. 
In section \ref{complete}
we introduce the fragmentation functions both for ordinary QCD and in 
the supersymmetric generalization proposed in \cite{bk}. Our conclusions 
are presented in section \ref{concl}.

\section{An estimate of the effect}\label{simple}

The SH particles we are interested in are quasistable, which means that 
their lifetime $\tau_X$ is much larger than the age $t_0$ of the 
Universe. If $n_0$ is the average density of SH particles with mass $m_X$ 
in the halo, the decay rate can be estimated as $\dot n_X^{halo}
\approx n_0/\tau_X$.
The decay of an X-particle typically results in the production
of a quark-antiquark pair 
$$
X \to q \bar q
$$
and each quark produces a jet of hadrons (mainly Nucleons ($N$) and
pions ($\pi$)) with total energy in the jet $E_{jet}=m_X/2$. 
The hadron spectrum is given by the fragmentation function 
$W_i(x,m_X)$ where $i=N,\pi$ and $x=E_i/E_{jet}=2E_i/m_X$. Many 
discussions can be found in the recent literature about approximations
to the fragmentation function and extrapolations of the low energy 
phenomenological expressions to the extremely large center of
mass energy involved in the production of UHECR by the decay of 
X-particles (for a recent review see \cite{sigl}; a numerical
approach involving the use of a QCD event generator was
used in \cite{sarkar}). We address  
this problem in the next section, while here we shall use
a very simple approximation proposed in \cite{hill,hsw} and valid for
$x\ll 1$. In this approach the spectrum of the pions, which represent most 
of the content of the jet, is given by
\begin{equation}
\frac{dN_{\pi}}{dx}=\frac{15}{16} x^{-3/2} (1-x)^2 \approx 
\frac{15}{16} x^{-3/2}~~~x\ll 1.
\label{eq:had}
\end{equation}
The spectrum of the nucleons has the same form as in eq. 
(\ref{eq:had}) but the normalization is such that the ratio $N/\pi$ 
in the jet is $\sim 0.05$ as suggested by the decay $Z_0\to hadrons$ 
at LEP. Moreover each pion family will take approximately $1/3$ of the
total pion content in the jet.\par
The small $N/\pi$ ratio at the production implies that on galactic distance
scales the flux of UHECR produced by the decay of X-particles
is dominated by the gamma rays produced in the decay of neutral pions.
Therefore, in the following we shall estimate the rate $\dot n_X^{halo}$ of
decay, requiring that the flux of gamma rays equals the flux of
UHECR at some energy.
For the sake of simplicity we shall also assume here that the SH
particles are distributed homogeneously in the halo whose size is $R_H$.
The flux of gamma rays with energy $E_\gamma$ at the Earth is then
easily derived to be
\begin{equation}
I_\gamma^{UHE}(E_{UHE})=2\times \frac{1}{3} \dot n_X^{halo} \frac{R_H}{4\pi}
\int_{E_{UHE}}^{E_{jet}}
dE_\pi \frac{dN_\pi}{dE_\pi} \frac{2}{E_\pi}~=~
\frac{5}{6} \dot n_X^{halo} \frac{R_H}{4\pi}
\left(\frac{m_X}{2}\right)^{1/2} E_{UHE}^{-3/2},
\label{eq:g_flux}
\end{equation}
where the factor $2$ in front takes into account the two jets and 
the factor $1/3$ accounts for the $\pi^0$'s only. Moreover we
used $dN_\pi/dE_\pi=(2/m_X)dN_\pi/dx$.\par
The comparison of our prediction with the observed flux of UHECR at $E=10^{10}$
GeV immediately gives
\begin{equation}
\dot n_X^{halo} \approx 1.8\cdot 10^{-42} \left(\frac{m_X}{10^{14}GeV}\right)^{-1/2}
\left(\frac{R_H}{100kpc}\right) cm^{-3} s^{-1}.
\label{eq:norma}
\end{equation}

A fraction $2/3$ of the energy in the jet is transformed into charged pions
which contribute an electron flux through production and decay of muons.
The density spectrum of these electrons can be calculated in the standard way
(muon decay) and gives
\begin{equation}
q_e(E_e)=\frac{40}{63} \dot n_X^{halo} \frac{m_\pi^3-m_\mu^3}{m_\pi^2-m_\mu^2}
\frac{1}{m_\pi}\left(\frac{m_X}{2}\right)^{1/2} E_e^{-3/2}.
\end{equation}
The ultra high energy electrons from pion decay are 
produced in the magnetic field of our galaxy, whose average 
strength will be denoted here by $B_\mu$ (in $\mu G$). The synchrotron
emission of these electrons will typically result in the production
of photons with energy 
\begin{equation}
E_\gamma = 1.5\cdot 10^3 B_\mu  \left(\frac{E_e}{10^{10}GeV}\right)^2 ~ GeV.
\label{eq:ensyn}
\end{equation}
Electrons with energy $E_e\geq 10^{17}$ eV will then radiate photons with 
energy $E_\gamma\geq 0.1$ GeV. Here we estimate the flux of these
photons. The synchrotron emission is the dominant channel of energy
losses at the energies of interest, and the time for losses is so small
that all the electron energy is very rapidly radiated in the form of
gamma rays, with flux
\begin{equation}
I_\gamma(E_\gamma)=\frac{R_H}{4\pi} q_e(E_e) \frac{E_e}{E_\gamma} 
\frac{dE_e}{dE_\gamma}=
1.2\cdot 10^{33} \left(\frac{m_X}{10^{14}GeV}\right)^{1/2} \dot n_X^{halo} B_\mu^{-1/4} 
\left(\frac{R_H}{100kpc}\right) E_\gamma^{-7/4} 
cm^{-2} GeV^{-1} s^{-1} sr^{-1}
\label{eq:gamma}
\end{equation}
where $\frac{dE_e}{dE_\gamma}$ has been derived from eq. (\ref{eq:ensyn}).
The normalized flux of gamma rays can now be calculated using 
eq. (\ref{eq:norma}) which gives the following very interesting result
\begin{equation}
I_\gamma(E_\gamma)=2.1\cdot 10^{-9} B_\mu^{-1/4} E_\gamma^{-7/4} ~ 
cm^{-2} GeV^{-1} s^{-1} sr^{-1}
\label{eq:bench}
\end{equation}
independent on the size of the halo $R_H$ and the mass of the X-particles 
$m_X$.\par
The result is also weakly dependent on the average magnetic field $B_\mu$.
For $B_\mu\sim 0.1-1~\mu G$ the flux derived from eq. (\ref{eq:bench}) turns
out to be detectable by EGRET in the energy range $100$ MeV$\leq 
E_\gamma\leq $ few GeV.

The inverse dependence of $I_\gamma$ on $B_\mu$ can be easily explained
on the basis of the electron spectrum resulting from the fragmentation 
function: such spectrum contains most of the energy at high electron energies.
Therefore, at fixed $E_\gamma$ the larger electron energies (correspondent
to smaller magnetic fields) contribute more, as far as the required electron 
energies are below $m_X/2$.\par
The result quoted in eq. (\ref{eq:bench}) changes appreciably as a 
function of the energy where SH relic particles provide the main 
contribution to the UHECR flux. This is more evident 
if we combine eqs. (\ref{eq:g_flux}) and (\ref{eq:gamma}) to give:
\begin{equation}
I_\gamma(E_\gamma)=8.2\cdot 10^3 I_\gamma^{UHE}(E_{UHE}) E_{UHE}^{3/2} 
B_\mu^{-1/4} E_\gamma^{-7/4},
\label{eq:normalizza}
\end{equation}
where now $E_{UHE}$ is the energy at which we normalize the UHECR flux 
$I_{\gamma}^{UHE}$ (measured in units of $cm^{-2} s^{-1} sr^{-1} GeV^{-1}$).
From eq. (\ref{eq:normalizza}) it results that if the normalization is
carried out at $5\cdot 10^{10}$ GeV, then the numerical factor in 
eq. (\ref{eq:bench}) is $\sim 18$ times smaller. In this case, for 
$B_\mu \sim 1$ the flux drops below the 
detectability level of EGRET, though it remains still in the detectability
range of
next generation satellites (e.g. GLAST). 

Let us come now to another interesting point: following the approach used
in \cite{bk} and \cite{bbv}, the rate of decay of SH particles in the halo
may be used to normalize the same quantity in the extragalactic space,
through the relation
\begin{equation}
\frac{\dot n_X^{halo}}{\dot n_{eg}} = \frac{\rho_{CDM}^{halo}}
{\Omega_{CDM}\rho_{cr}}
\label{eq:extra}
\end{equation}
where $\dot n_{eg}$ is now the rate of X particle decay in the extragalactic
space, $\rho_{CDM}^{halo}=0.3$ GeV$cm^{-3}$ is the energy density of CDM 
in the halo, $\Omega_{CDM}$ is the fraction of CDM and $\rho_{cr}$ is the
critical density of the Universe. We shall adopt here $\Omega_{CDM} h^2=0.2$.
If we call $B_\mu^{eg}$ (in $\mu G$) the extragalactic magnetic field, the
flux of gamma rays due to the synchrotron emission of UHEE can be calculated
similarly to eq. (\ref{eq:gamma}) with $R_H$ substituted by $ct_0$, where
$t_0=2.06\cdot 10^{17} h^{-1} s$ is the age of the Universe and $h$ is the
dimensionless Hubble constant. Of course this treatment is valid
as far as gamma rays are not absorbed on scales much smaller than $ct_0$, 
which limits this estimate to $E_\gamma< 500-1000$ GeV. In this assumption
the ratio of gamma rays with energy $E_\gamma$ produced in the extragalactic
space relative to the ones produced in the halo is given by
\begin{equation}
\frac{I_\gamma^{eg}(E_\gamma)}{I_\gamma(E_\gamma)} = 0.23
\left(\frac{R_H}{100 kpc}\right)^{-1} \left(\frac{\Omega_{CDM} h^2}
{0.2}\right) \left( \frac{B_\mu}{B_\mu^{eg}}\right)^{1/4},
\label{eq:gal_ex}
\end{equation}
where we used $h=0.6$.
For extragalactic magnetic fields of order $10^{-9}-10^{-10}$ Gauss, 
and a typical galactic field of order of $\mu G$, the two gamma ray fluxes 
are comparable. Unfortunately only upper limits are currently available on the
extragalactic magnetic field. Therefore in the following we shall not 
consider the extragalactic flux in any more detail and we shall concentrate
our attention on the gamma ray flux from the halo.

\section{A detailed calculation}\label{complete}

The simple calculations presented in the previous section suggest that 
the synchrotron emission of UHEE in the galactic
magnetic field can produce an interesting flux in high energy gamma 
rays. The rate of production of electrons is normalized by requiring that
the gamma rays generated in the decay of neutral pions saturate the
flux of UHECR above some energy. In this section
we directly use as normalization point $5\cdot 10^{19}$ eV, rather 
than the $10^{19}$ eV used before, mainly inspired by two factors:
1) using more realistic fragmentation functions and normalizing our
gamma ray fluxes at $10^{19}$ eV it is easy to overproduce the UHECR 
at higher energy; 2) in the same conditions the flux of primary
gamma rays can easily exceed the CASA-MIA limits at $\sim 10^8$ GeV.
As a consequence of this more conservative approach, the fluxes that 
will be derived here will be appreciably smaller than the ones 
derived in the previous bentchmark calculation, and can be considered as
a lower limit prediction.

As first pointed out in \cite{bkv} and \cite{bbv} the UHECR
flux in this halo model is the generation spectrum resulting from the
initial fragmentation process, hence it is dominated by gamma rays
with respect to nucleons.\par
In this section we assume, following \cite{bbv}, that 
the rate of particle
production is parametrized as the dark matter distribution \cite{kkbp}:
\begin{equation}
\dot n_X^h(R)=\frac{\dot n_0^h}{(R/r_0)^\gamma \left[1+(R/r_0)^\alpha
\right]^{(\beta-\gamma)/\alpha}}
\label{eq:dm}
\end{equation}
where $\dot n_0^h$ is the normalization rate, $R$ is the distance to the 
galactic center and $r_0$ is a distance scale between $5$ and $10$ kpc.
The parameters $\alpha,\beta,\gamma=(2,2,0)$ correspond to an isothermal
profile, while $\alpha,\beta,\gamma=(2,3,0.2)$ give the best fit to 
observational 
data \cite{kkbp} and finally $\alpha,\beta,\gamma=(1,3,1)$ are obtained in the
numerical simulation of ref. \cite{nfw}.\par
The calculation of the spectrum of hadrons produced in a single X-particle
decay is given by the fragmentation function for the process $q\to hadrons$.
The fragmentation functions we use here are the MLLA limiting spectrum 
of ordinary QCD \cite{dok} and its supersymmetric generalization given in 
\cite{bk} (hereafter SUSY-QCD). 
In the first case the fraction of pions produced
in a single event is taken as $f_\pi\approx 1$, while in the second
case we take, according with \cite{bk}, 
$f_\pi\approx 0.5$. The pion fragmentation
functions, indicated by $W_\pi(x,m_x)$ are then normalized as
\begin{equation}
\int_0^1 dx x W_\pi(x,m_x)=f_\pi
\label{eq:normalizzaff}
\end{equation}
where $x=2E_\pi/m_x$. Approximately $1/3$ of $f_\pi$ is carried away
by neutral pions while a fraction $2/3$ goes into charged pions. Each 
X-particle will produce two jets with energy $m_X/2$ each.\par

The choice of the fragmentation function and its normalization are
delicate points and deserve some additional comments: both the MLLA
fragmentation functions considered here (QCD and 
SUSY-QCD) are still valid for $x\ll 1$, though they represent an 
improvement with respect to the simple approximation in eq. (\ref{eq:had}). 
This implies a lack of consistency in the way we normalize the 
fragmentation function through eq. (\ref{eq:normalizzaff}) since the
integral extends to a range of values of $x$ where the fragmantation
function is not accurate. Recently this problem has been treated more
carefully in \cite{sarkar} and several new interesting results have
been obtained. In \cite{sarkar} the HERWIG QCD event generator was
used to investigate the structure of the hadronic cascade at large
values of $m_X$. In this approach the normalization of the fragmentation
function can be carried out self consistently. More interestingly however,
the authors of \cite{sarkar} also find appreciable differences between 
the fragmentation functions produced by their montecarlo tecnique and
the functions generally used in these calculations as well as in the
present work. The differences, as could be expected, vanish for 
$m_X\leq 10^3$ GeV, where the MLLA approximation is known to work well.
A striking difference between the montecarlo approach and the MLLA
approximation is that the fragmentation functions for protons and pions 
(and therefore neutrinos and gamma rays) have different shapes for 
large values of $x$ (while we assumed here that the shape of the two
functions is the same but with a different normalization). Moreover,
for $0.2\leq x\leq 0.4$ the authors of \cite{sarkar} seem to obtain
equal abundancies of nucleons and gamma rays, which is a very different 
result from the usual assumption that the gamma rays are
a factor $\sim 10$ more abundant than nucleons in the hadronic cascade.
The combination of different normalization and different shape of the
fragmentation functions suggested in \cite{sarkar} can affect the
conclusions presented in this paper and more in general the calculations
of particle fluxes in top-down models of UHECR. Unfortunately, since
the results in \cite{sarkar} are purely numerical it is not easy 
to envision in what
direction the results presented here would change, therefore, we assume
here that the fragmentation process is well represented in the MLLA 
approach introduced above, stressing that a detailed study of the 
problem would certainly be useful.

Since the electrons radiate through synchrotron emission, a model
for the galactic magnetic field is also needed. We parametrize the
magnetic field in the form 
\begin{equation}
B(r,z)=B_0(r) exp(-z/z_0)
\label{eq:magn}
\end{equation}
where $B_0(r)=3(r_\odot/r)\mu G$ when the radial distance $r$ in the disc
is larger than $4$ kpc and $B_0(r)=$const for $r<4$ kpc 
(e.g. \cite{stanev}). The parameter 
$z_0$ is fixed at $0.5$ kpc and $r_\odot=8.5$ kpc is the distance of
the sun relative to the galactic center. $z$ is the height above
the galactic disc.\par
According to \cite{bbv} the flux of gamma rays with energy $E_\gamma$ 
produced in the $\pi^0$ decay is given by
\begin{equation}
I_\gamma^{UHE}(E_\gamma)=\frac{W_\gamma(E_\gamma)}{4\pi}
\int_{-1}^{1} dcos\theta |cos \theta| \int_0^{r_{max}(\theta)}
dr \dot n_X^h(R)
\label{eq:gamma_det}
\end{equation}
where $r_{max}(\theta)$ is easily calculated to be
\begin{equation}
r_{max}(\theta)=r_\odot cos \theta + \sqrt{R_h^2-r_\odot^2 sin^2 \theta}
\label{eq:rmax}
\end{equation}
and $R_h$ is the size of the halo. The angle $\theta$ is measured 
with respect to 
the direction of the galactic center.
The gamma ray spectrum at the production is given by \cite{bbv}
\begin{equation}
W_\gamma(E_\gamma)=\frac{4}{m_X}\int_{2E_{\gamma}/m_X}^{1} 
\frac{dx}{x} W_{\pi^0}(x,m_X)
\label{eq:Wg}
\end{equation}
with $W_{\pi^0}(x,m_X)=(1/3)W_{\pi}(x,m_X)$.\par
The comparison of the observed UHECR flux at $E=5\cdot 10^{10}$ GeV with the 
flux of gamma rays from eqs. (\ref{eq:gamma_det}-\ref{eq:Wg}) at the same
energy gives the normalizing value of $\dot n_0^h$ for each specific
choice of the values of the parameters. We plot our fluxes of UHE 
gamma rays at the normalization point in Fig. 1.

The spectrum of UHEE is the convolution of
the pion and muon spectra and is given by
\begin{equation}
W_e(E_e)=12 E_e^2 \frac{m_\pi^2}{m_\pi^2-m_\mu^2} \frac{2}{m_X}
\int_{E_e}^{m_X/2} dE_\mu \frac{1}{E_\mu^3}
\left[ 1-\frac{E_e}{E_\mu}\right] 
\int_{2E_\mu/m_X}^{2r^2E_\mu/m_X}  \frac{dx}{x} W_{\pi^{\pm}}(x,m_X),
\label{eq:elec}
\end{equation}
where $r=m_\pi/m_\mu$. 
The spectrum of electrons produced by a single muon is peaked at
$E_e\approx (2/3)E_\mu$. In the simple assumption that all electrons
are produced with this peak energy the above electron spectrum becomes 
\begin{equation}
W_e(E_e)=\frac{3}{m_X}\frac{m_\pi^2}{m_\pi^2-m_\mu^2}
\int_{3E_e/m_X}^{3r^2E_e/m_X} \frac{dx}{x} W_{\pi^{\pm}}(x,m_X).
\label{eq:simple}
\end{equation}
As pointed out in the previous section the UHEE radiate by synchrotron
emission in the galactic magnetic field in the gamma ray energy range,
according with eq. (\ref{eq:ensyn}). The emissivity per unit volume in
the form of gamma rays with energy $E_\gamma$ at the position at distance
$R$ from the galactic center can be written as
\begin{equation}
q_\gamma(E_\gamma,r,z)=\dot n_X^h(R) W_e(E_e) \frac{E_e}{E_\gamma}
\frac{dE_e}{dE_\gamma}
\end{equation}
where $E_e$ and $E_\gamma$ are related through eq. (\ref{eq:ensyn}) but
the magnetic field is now a function of the position in the galaxy. Hence
the same $E_\gamma$ will correspond to different $E_e$ in different 
positions in the galaxy.\par
Since the magnetic field is symmetric with respect to the disc while the 
dark matter profile is spherically symmetric, the integration over 
lines of sight requires the explicit integration over 
both $\theta$ and $\phi$, so
that the flux of gamma rays at the Earth, per unit time, surface,
energy and solid angle is 
\begin{equation}
I_\gamma(E_\gamma)=\frac{1}{4\pi}
\int_{0}^{2\pi} d\phi \int_{-1}^1 dcos\theta
~|cos\theta| \int_0^{r_{max}(\theta)} dr q_\gamma(E_\gamma,r,z),
\label{eq:I_syn}
\end{equation}
where $r_{max}(\theta)$ is again given by eq. (\ref{eq:rmax}).\par
The total observed gamma ray flux at energy $E_\gamma$, 
$I_\gamma^{tot}(E_\gamma)$, is the sum of the 
synchrotron gamma ray flux [eq. (\ref{eq:I_syn})] and the primary gamma ray
flux [eq. (\ref{eq:gamma_det})].

The results of the calculations are plotted in Figs. 2 and 3, where we 
used $r_0=10$ kpc and $z_0=0.5$ kpc. Although the calculation has been 
carried out 
for the three sets of values of the $\alpha,\beta,\gamma$ parameters,
the results are quite insensitive to the specific model of the halo
distribution of dark matter, and the related curves would be indistiguishable. 
Therefore we plotted there only the case 
$(\alpha,\beta,\gamma)=(2,2,0)$ for reference. 
In Fig. 2 we plotted $E_\gamma^{1.75} I_\gamma^{tot}(E_\gamma)$ 
in order to amplify
the spectral differences. The two thick lines refer to 
$m_X=10^{14}$ GeV, while the two thin lines are for $m_X=10^{13}$ GeV.
The solid curves are obtained using the SUSY-QCD fragmentation function
while the dashed lines are calculated with the ordinary QCD fragmentation
function. For comparison we also plotted by a dash-dotted line the 
extragalactic diffuse gamma ray background \cite{EGRET}. 
All our fluxes are below this limit in the 
energy range accessible to current experiments. The points with arrows
are the upper limits on the gamma ray flux from the HEGRA \cite{hegra},
Utah-Michigan \cite{utah}, EAS-TOP \cite{eastop} and CASA-MIA \cite{casa}
experiments.\par
Clearly the stronger constraints on the gamma ray flux from the halo
are imposed by the CASA-MIA and EAS-TOP experiments. In most of the 
cases considered 
in this paper the fluxes are below these limits, with the exception of
the case $m_X=10^{13}$ GeV with SUSY-QCD fragmentation function
(thin solid line). Note that in this case the overproduction of gamma 
rays at $\sim 10^7-10^8$ GeV is mainly due to the primary gamma rays
from $\pi^0$ decay, rather than to the synchrotron flux.

In Fig. 3 the gamma ray fluxes have been blown up in the energy range
between $100$ MeV and $10^4$ GeV (lines labelled as before) and 
compared with the sensitivity of
some of the present or planned gamma ray experiments 
in the same energy range (e.g. EGRET, GLAST, MILAGRO).

It could be particularly interesting to look at the gamma ray signal
in the region above $\sim 500$ GeV, where the contamination due to the
isotropic diffuse gamma ray background, believed to be of extragalactic 
origin, should be
reduced due to absorption of gamma rays on the infrared background. 
Our calculations show an integrated flux above $500$ GeV at the level 
of $\sim 10^{-12}$ photons $cm^{-2}s^{-1}sr^{-1}$ in our best case
scenario.

There is clearly a dependence of the results on $m_X$, introduced by 
the fragmentation functions,
more complicated than the simple power law adopted in section \ref{simple}. 
The difference is particularly evident on the tail of the spectra where the 
highest energy electrons (with energy close to $\sim m_X/2$) are probed.

We also checked that our predictions do not depend appreciably on the typical 
scale height $z_0$ of the magnetic field (eq. (\ref{eq:magn})).\par

A few comments about the results presented in the figures are needed: 
the factors $4\pi$ in eqs. (\ref{eq:gamma_det}) and (\ref{eq:I_syn})
scale the fluxes to the unit solid angle; this is the correct procedure
only for an isotropic distribution of the arrival directions of the 
gamma rays. Due to the anisotropic distribution of the dark matter halo
and of the magnetic field of the Galaxy as seen from the sun, neither 
the synchrotron flux
nor the primary gamma ray flux are isotropic. In fact our preliminary 
results of the anisotropy for the synchrotron gamma rays \cite{blasi} shows a 
pattern of arrival directions which is  quite different from the one of primary
gamma rays, calculated in \cite{bbv} and \cite{russi}. In this sense the
fluxes per unit solid angle plotted in the figures must be intended as 
indicative of the `real' fluxes, but more detailed calculations are
definitely needed in this direction. For the same reason, summing up the
contribution of primary and synchrotron gamma rays as done above is not
self consistent for fluxes anisotropic in different ways. 
A peculiar finding in the anisotropy
pattern could be important to discriminate the synchrotron gamma rays
produced by SH particles from other background components and could 
improve the possibility to detect the signal discussed here.

\section{Conclusions}\label{concl}

We calculated the flux of gamma radiation produced through synchrotron emission
in the galactic magnetic field from ultra-high energy electrons generated as
decay product of charged pions in the hadronic cascade coming from the
decay of X-particles, when they cluster in the galactic halo in the
form of SH ($m_X\sim 10^{13}-10^{14}$ GeV) quasistable particles. 

If the magnetic field in the galactic halo is
$B_{halo}\sim 0.1-1\mu G$, UHEE with energy between $10^{17}$ eV 
and $\sim m_X/2$ very rapidly loose all their
energy in the form of gamma rays mostly with energies $E_\gamma\leq 10^6-
10^8$ GeV. The resulting diffuse gamma ray emission above $100$ MeV is
of the order of $10^{-8}$ photons $cm^{-2} s^{-1} sr^{-1}$ in the best 
case scenario. 
This result is not strongly dependent on the mass $m_X$ of the
SH particles (though some dependence is introduced due to the 
fragmentation functions used here) nor on the geometry of the 
galactic dark matter 
halo, but depends critically on the energy where the contribution of SH 
relic particles to UHECR becomes dominant (section \ref{complete}).
The shape of the fragmentation functions at very high energy 
can affect the result in several ways, and an accurate procedure of
extrapolation of the fragmentation functions tested in accelerator experiments
up to extremely large values of $m_X$ is needed, as stressed in \cite{sarkar}.
In fact in \cite{sarkar} appreciable differences were found respect to 
the MLLA approximation used here, when the fragmentation functions
are determined by using the HERWIG QCD event generator.

In section \ref{simple} we also estimated the extragalactic 
flux of gamma rays due to the same
physical process, and found that, for typical extragalactic magnetic 
fields of $\sim 10^{-10}-10^{-9}$ Gauss the extragalactic and the halo
fluxes are comparable, if SH particles provide a fraction 
$\Omega_{CDM}h^2=0.2$ of the CDM and the SH particles in the halo
explain the UEHCR flux (the galactic and extragalactic densities of
SH particles are related through eq. (\ref{eq:extra})).\par
It is interesting to note that the detection
of a residual gamma ray emission from the halo, after the subtraction of 
the contribution of $pp$ collisions, Inverse Compton scattering, electron
bremsstrahlung and of the extragalactic diffuse gamma ray background was 
recently claimed. This subtraction procedure is quite difficult and very 
model dependent; however the analysis in \cite{halo} and \cite{dark} 
show that the residual emission is at the level of $\sim 10^{-7}$ 
photons $cm^{-2} s^{-1} sr^{-1}$ above $1$ GeV with a spectrum 
$E_\gamma^{-\alpha}$ and $\alpha\approx 1.8$.
This flux, if confirmed, is larger than what is found 
in the present calculation with the normalization at the UHECR flux carried
out at $5\cdot10^{10}$ GeV, but the power index $\alpha\approx 1.8$ is close 
to $1.75$ in eq. (\ref{eq:bench}). In the more realistic calculations 
illustrated in section \ref{simple} the spectrum cannot be approximated 
by a unique 
power law but in the energy region between $100$ MeV and a few GeV the slope 
of the spectra in the log-log plot is around $1.8-1.9$.\par
The origin of the residual emission is still under debate and the 
possibility of an underestimate of the contribution of inverse Compton
scattering is an open possibility, so that the actual residual emission
(if any) could be smaller than what previously claimed.
In this prospective it is interesting to study the contribution of SH
particles in the halo to this emission as an open intriguing possibility
requiring further investigation, though probably only next generation 
gamma ray detectors can shed a new light on the problem.\par
A particularly interesting aspect is the anisotropy of the gamma ray 
emission expected in this model \cite{blasi}.
The anisotropy is the result of both
the asymmetric configuration of the magnetic field in the galaxy
as seen from the Earth, and the asymmetric position of the solar system 
in the galaxy (this also produces some anisotropy in the arrival 
directions of UHECR \cite{russi,bbv,bm}).\par
The upper limits reported in Fig. 2 are due to the HEGRA \cite{hegra},
Utah-Michigan \cite{utah}, EASTOP \cite{eastop} and CASA-MIA \cite{casa}
experiments. The stronger constraints to the model of SH particles
in the halo come from the CASA-MIA and EASTOP experiments:
the calculated fluxes are compatible with the experimental limits
with the exception of the case
obtained for $m_X=10^{13}$ GeV and the SUSY-QCD fragmentation function.
In this case the exceeding limit is mainly due to the primary gamma rays 
from $\pi^0$ decay rather that to the synchrotron gamma rays.\par

We propose that it could be of particular interest to study the gamma
ray emission from the halo in the energy range $E_\gamma\geq 500-1000$ GeV, 
for two reasons: a) the diffuse extragalactic gamma ray background is 
expected to be less pronounced in this energy range because of absorption
of gamma rays in the intergalactic medium on the infrared photon
background; b) the spectrum of the gamma radiation from SH relic particles
is predicted to have a spectrum $E_\gamma^{-\alpha}$ with $\alpha\approx
1.7-2$, substantially flatter than the gamma ray emission from $pp$
collisions (expected to have the same spectrum as the cosmic rays) which
should dominate over other mechanisms at these high energies.

\vskip 1cm
{\bf Aknowledgments}

The author is grateful to V.S. Berezinsky and A. Olinto for many 
useful discussions that helped understanding the 
proposed effect and for a critical reading of the manuscript, to 
R. Ong and C. Covault for discussions on the observational limits
and to G. Sigl for numerous comments. The research of P.B. is funded 
by INFN at the University of Chicago.

\newpage
{\bf Figure Captions}
\vskip 1cm

{\bf Fig. 1}: Flux of UHE gamma rays from $\pi^0$ decay for the 
case of SH particles in the halo. The thick lines are for 
$m_X=10^{14}$ GeV and the thin lines are for $m_X=10^{13}$ GeV. The 
solid lines are obtained with the SUSY-QCD fragmentation 
function \cite{bk}, while the dashed lines are for the ordinary QCD 
fragmentation function \cite{dok}. The fluxes are multiplied by
$E^3$. Data points are from ref. \cite{agasa}.
\vskip 1cm

{\bf Fig. 2}: Total flux of gamma rays as sum of the synchrotron 
emission of UHEE in the galactic magnetic field and the contribution
of the decay of neutral pions.
Lines are labelled as in Fig. 1. Fluxes are multiplied by $E_\gamma^{1.75}$
to emphasize the spectral differences among the curves.
Upper limits are from the HEGRA \cite{hegra}, Utah-Michigan \cite{utah},
EASTOP \cite{eastop} and CASA-MIA \cite{casa} experiments. The dash dotted
line is the extragalactic diffuse gamma ray background from EGRET \cite{EGRET}.

\vskip 1cm

{\bf Fig. 3}: Flux of gamma rays (multiplied by $E_\gamma$) for the same cases
as in Fig. 2 but limited to the energy range $100$ MeV $\leq E_\gamma
\leq 10^4$ GeV. Lines are labelled as in Fig. 1. The sensitivity
of the EGRET, GLAST and MILAGRO experiments are drawn.


\begin{references}

\bibitem{agasa}
M. Takeda et al., preprint astro-ph/9807193.

\bibitem{gzk}
K. Greisen, Phys. Rev. Lett. {\bf 16}, 748 (1966); G.T. Zatsepin and
V.A. Kuzmin, Sov. Phys. JETP Lett. {\bf 4}, 78 (1966).

\bibitem{kronberg}
P.P. Kronberg, Rep. Prog. Phys. {\bf 57}, 325 (1994).

\bibitem{bo}
P. Blasi, S. Burles and A. Olinto, Astrophys. J. Lett. {\bf 514}, L79 (1999).

\bibitem{sigl}
P. Bhattacharjee and G. Sigl, preprint astro-ph/9811011 (submitted to 
Phys. Rep.).

\bibitem{bkv}
V.S. Berezinsky, M. Kachelriess and A. Vilenkin, 
Phys. Rev. Lett. {\bf 79}, 4302 (1997).

\bibitem{sarkar}
M. Birkel and S. Sarkar, Astropart. Phys. {\bf 9}, 297 (1998). 

\bibitem{kuz}
V.A. Kuzmin, at the Workshop {\it Beyond the Desert}, Castle Ringberg, 
June 1997 (preprint astro-ph/9709187) and 
{\it International Workshop on Non Accelerator New Physics}, Dubna, july 1997.

\bibitem{bere}
V.S. Berezinsky, preprint astro-ph/9811268.

\bibitem{kolb1}
D.J.H. Chung, E.W. Kolb and A. Riotto, preprint astro-ph/9805473;
D.J.H. Chung, E.W. Kolb and A. Riotto, preprint astro-ph/9809453.

\bibitem{russi}
S.L. Dubovsky and P.G. Tinyakov, preprint hep-ph/9802382.

\bibitem{bbv}
V.S. Berezinsky, P. Blasi and A. Vilenkin, Phys. Rev. {\bf D58}, 103515 (1998).

\bibitem{bm}
V.S. Berezinsky and A. Mikhailov, preprint astro-ph/9810277.

\bibitem{bk}
V.S. Berezinsky and M. Kachelriess, Phys. Lett. {\bf B434}, 61 (1998).

\bibitem{hill}
C.T. Hill, Nucl. Phys. {\bf B224}, 469 (1983).

\bibitem{hsw}
C.T. Hill, D.N. Schramm and T.P. Walker, Phys. Rev. {\bf D36}, 1007 (1987).

\bibitem{kkbp}
A.V. Kravtsov, A.K. Klypin, J.S. Bullock and J.R. Primack, Astrophys. J.
{\bf 502}, 48 (1998).

\bibitem{nfw}
J.F. Navarro, C.S. Frenk and S.D.M. White, Astrophys. J. {\bf 462}, 563 (1996).

\bibitem{dok}
Yu.L. Dokshitzer, V.A. Khose, A.H. Mueller and S.I. Troyan, {\it Basics 
of Perturbative QCD} (Editions Fronti\`eres, Gif-sur-Yvette, France, 1991).

\bibitem{stanev}
T. Stanev, Astrophys. J. {\bf 479}, 290 (1997).
 
\bibitem{hegra}
A. Karle et al., Phys. Lett. {\bf B347}, 161 (1995).

\bibitem{utah}
J. Matthews et al., Astrophys. J. {\bf 375}, 202 (1991).

\bibitem{eastop}
M. Aglietta et al., Astropart. Phys. {\bf 6}, 71 (1996).

\bibitem{casa}
M.C. Chantell et al., Phys. Rev. Lett. {\bf 79}, 1805 (1997).

\bibitem{halo}
D.D. Dixon et al., New Astronomy {\bf 3}, 539 (1998).

\bibitem{dark}
R. Chary and E.L. Wright, preprint astro-ph/9811324.

\bibitem{blasi}
P. Blasi and A. Olinto, in preparation.

\bibitem{EGRET}
A. Chen, J. Dwyer and P. Kaaret, Astrophys. J. {\bf 463}, 169 (1996);
P. Sreekumar et al., Astroph. J. {\bf 494}, 523 (1998).


\end{references}
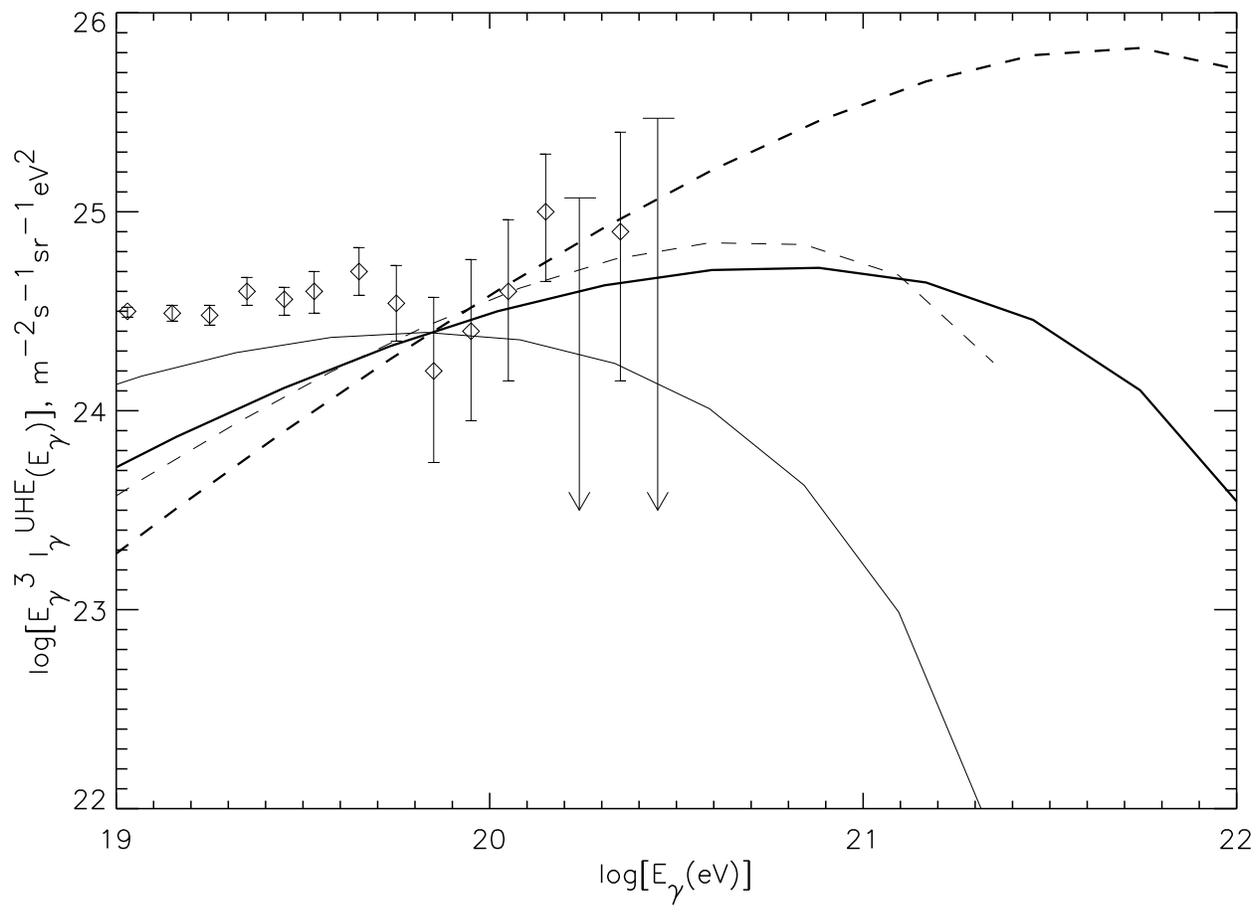
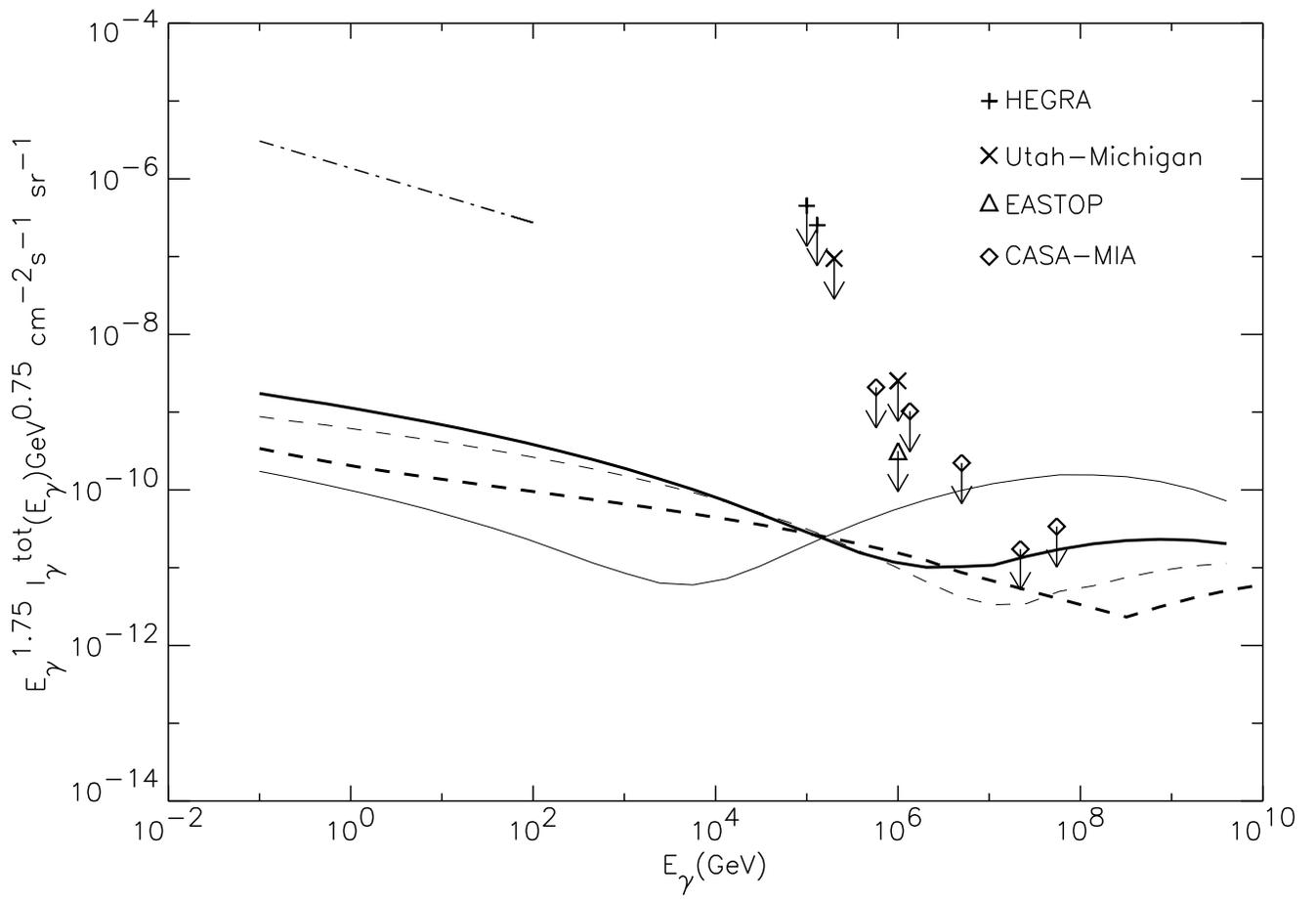
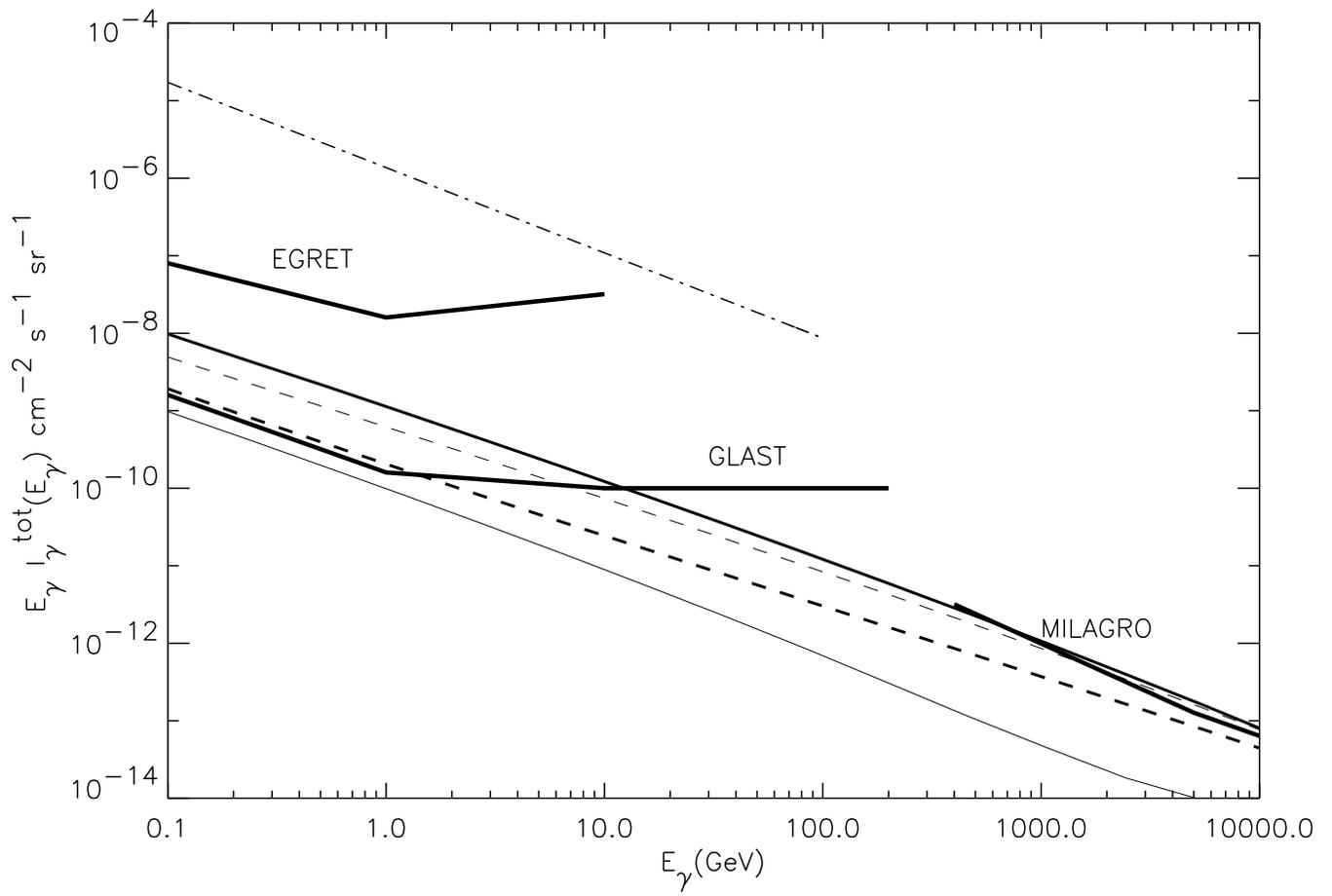
\end{document}